# Specific Features of Magnetic Phase Diagram of an MnSi Helimagnet[1]

V. N. Narozhnyi* and V. N. Krasnorussky

*Vereshchagin Institute for High Pressure Physics, Russian Academy of Sciences, Troitsk, Moscow oblast, 142190 Russia*
*e-mail: narozhnyivn@gmail.com*

**Abstract**—Magnetization curves of MnSi single crystals have been measured in a range of temperatures $T = 5.5–35$ K and magnetic field strengths $H \leq 11$ kOe for **H** || [1 1 1], [0 0 1], and [1 1 0]. Special attention has been paid to the temperature interval near $T_N = 28.8$ K, where MnSi exhibits a transition to the state with a long-period helical magnetic structure. Some new features in the magnetic behavior of MnSi have been found. In particular, in an intermediate temperature region above the transition ($28.8$ K $= T_N \leq T < 31.5$ K), the $dM(H)/dH$ curves exhibit anomalies that are not characteristic of the typical paramagnetic state. It is established that the line of the characteristic field $H^*(T)$ of this anomaly is a natural extrapolation of the temperature dependence of the field of the transition from a conical phase to an induced ferromagnetic phase observed at $T < T_N$. It is concluded that the properties of MnSi in the indicated intermediate region are related to and governed by those of the conical phase (rather than of the $A$ phase). Based on these data, magnetic phase diagrams of MnSi for **H** || [1 1 1], [0 0 1], and [1 1 0] are plotted and compared to diagrams obtained earlier by other methods.

## 1. INTRODUCTION

The intermetallic compound MnSi crystallizes in a cubic (B20 type) lattice without a center of inversion. It is commonly accepted that the magnetic properties of MnSi predominantly have an itinerant (band) character. The absence of a center of inversion permits Dzyaloshinskii–Moriya-type interaction in this system (in addition to the usual exchange), which in turn leads to the formation of a long-period helical magnetic structure (at $T < T_N = 28.8$ K) with the orientation of helix axes determined by the weaker anisotropic exchange interaction (see, e.g., [1]).

A specific feature of MnSi is the existence of an intermediate region adjacent to the magnetic phase transition from higher temperatures. In this region, the heat capacity, magnetic susceptibility, temperature coefficient of electric resistance, etc., exhibit anomalous behavior (see, e.g., [2]). The nature of this behavior in the properties of MnSi is still not completely clear.

## 2. EXPERIMENTAL TECHNIQUES

Systematic measurements of the $M(H)$ static magnetization curves of MnSi single crystals have been performed in a range of temperatures $T = 5.5–35$ K and magnetic field strengths up to $H = 11$ kOe for **H** || [1 1 1], [0 0 1], and [1 1 0]. Special attention has been devoted to temperature interval near $T_N = 28.8$ K, where peaks in the heat capacity, temperature coefficient of thermal expansion, and magnetic susceptibility have been observed in some earlier investigations. The experiments were performed with MnSi single crystals grown by the Bridgman technique.

The magnetization was measured using a vibrating-sample magnetometer (Lake Shore Cryotronix Inc.) for the three orientations of the external magnetic field indicated above. The transverse configuration of the magnetic field in the installation and the possibility of rotating the axis of sample suspension allowed the magnetization to be measured for all three orientations of the field for one sample setting in the holder. The sample had a cubic shape with an edge length of about 3 mm and faces oriented perpendicular to the [1 1 0], [1 $\bar{1}$ 0], and [0 0 1] directions. High-precision $M(H)$ measurements allowed the differential magnetic susceptibility $dM(H)/dH$ to be reliably determined by numerical differentiation and, in turn, some new features in the magnetic behavior of MnSi to be revealed in various temperature intervals and magnetic fields.

## 3. RESULTS AND DISCUSSION

In the temperature interval of $5.5$ K $\leq T < 27.0$ K, the $M(H)$ dependence is close to linear in fields up to $H \approx 4–6$ kOe. Figure 1 presents examples of magnetization $M(H)$ and the corresponding derivative $dM(H)/dH$ curves of MnSi determined for **H** || [1 1 1] in three temperature intervals. A further increase in

---

[1] The article is based on a preliminary report delivered at the 36th Conference on Low-Temperature Physics (St. Petersburg, July 2–6, 2012).

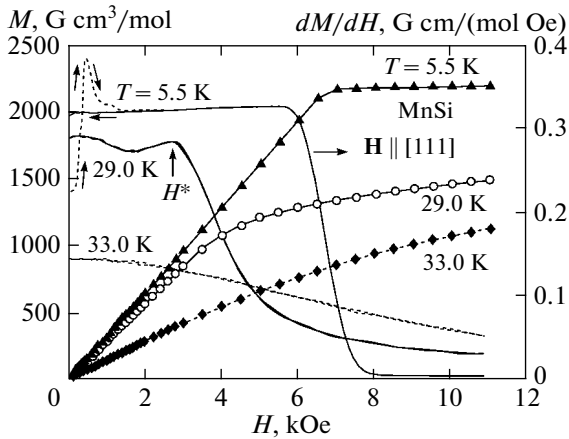

**Fig. 1.** Typical plots of magnetization $M$ and differential magnetic susceptibility $dM(H)/dH$ vs. magnetic field $H$ for MnSi at $\mathbf{H} \parallel [1\,1\,1]$ and $T = 5.5$, 29.0, and 33.0 K. The arrows near the $dM(H)/dH$ curve for $T = 5.5$ K indicate the direction of field variation; curves connecting $M(H)$ points are drawn for better illustration.

the magnetic field led to the appearance of a rather sharp bending on the $M(H)$ curves. This behavior of the magnetization curves at $T < T_N$ is well known and attributed to the transition of the sample into a field-induced ferromagnetic state at sufficiently high fields (see, e.g., [3]).

Our investigation has revealed an additional feature on both $M(H)$ and $dM(H)/dH$ curves of MnSi, which was observed at temperatures within $5.5 \text{ K} \leq T < 28.8$ K for various field orientations with $H$ values in the range from 80 Oe to 1.3 kOe (see the results of magnetization measurements at $T = 5.5$ K in Fig. 1 and the phase diagrams for three $\mathbf{H}$ orientations plotted in Figs. 2–4 based on the results of our measurements). It should be noted that anomalies in the properties of MnSi in this range of fields have also been observed previously, when this compound was studied by other methods (e.g., using ultrasound absorption, neutron scattering, and ac magnetic susceptibility measurements).

Our investigation showed that the position of this specific feature significantly depends on both the direction of applied magnetic field and temperature (see Figs. 2–4). Note that this anomaly for $\mathbf{H} \parallel [1\,1\,0]$ falls in between its positions for $\mathbf{H} \parallel [1\,1\,1]$ and $\mathbf{H} \parallel [0\,0\,1]$. In addition, we have observed a significant irreversibility for the variation of $M(H)$ in the region of this anomaly for $\mathbf{H} \parallel [1\,1\,1]$ (see the $dM(H)/dH$ curve for $T = 5.5$ K in Fig. 1). This irreversibility for $M(H)$ at $\mathbf{H} \parallel [1\,1\,1]$ is retained for temperatures of up to $T = 27.4$ K. At the same time, in the vicinity of the phase transition ($27.6 \text{ K} \leq T \leq 28.8$ K), the $M(H)$ curve becomes practically reversible (this region is indicated by open square symbols in Fig. 2).

The $M(H)$ curves for the two other field directions are almost reversible for all temperatures. The corresponding positions of anomalies on the $M(H)$ curves near $\mathbf{H} \approx 1$ kOe are plotted with open circles in Figs. 3 and 4. Some irreversibility that is still present in both $M(H)$ and $dM(H)/dH$ for these field directions can be characterized by the width of the "hysteresis" loop observed during the sequential increase and decrease in the field strength. This hysteresis width does not exceed 200 Oe. It is also important that, as the field decreases to zero, the $dM(H)/dH$ value is almost completely restored at the initial level (in contrast to the case of $\mathbf{H} \parallel [1\,1\,1]$). For $\mathbf{H} \parallel [1\,1\,0]$, the peak on the $dM(H)/dH$ curve measured near $T_N$ exhibited splitting into two components. The field dependence of the additional component is plotted by open triangles in Fig. 4. This behavior can be related to instability of the helical structure for the given direction of **H**, at least in the vicinity of $T_N$.

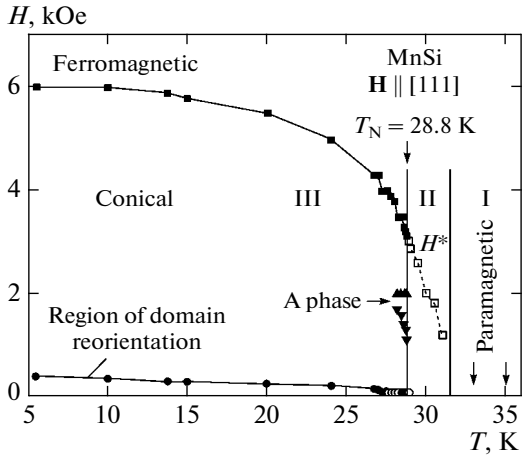

**Fig. 2.** Magnetic phase diagram of MnSi for $\mathbf{H} \parallel [1\,1\,1]$. Open squares connected by dashed line $H^*(T)$ in the interval of $28.8 \text{ K} \leq T \leq 31.0$ K correspond to the positions of specific features on $dM(H)/dH$ curves at various temperatures. The arrows at $T = 33.0$ and $35.0$ K indicate the temperatures at which no anomalies have been found on the $dM(H)/dH$ curves at fields down to the minimum studied. Open symbols in the range of $27.6 \text{ K} \leq T \leq 28.8$ K correspond to the positions of an anomaly related to the reorientation of magnetic domains and almost reversible behavior of magnetization during variation of the applied field. Black symbols show the positions of anomalies on irreversible $M(H)$ curves.

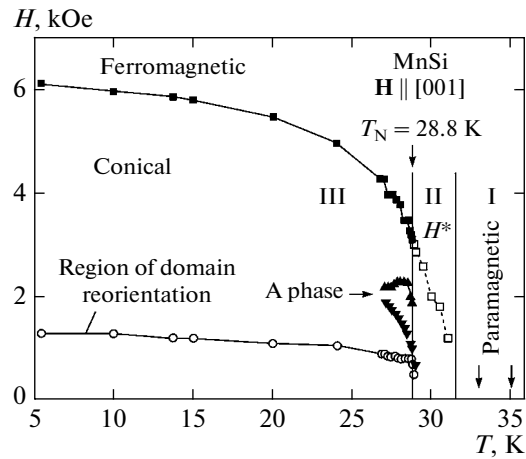

**Fig. 3.** Magnetic phase diagram of MnSi for $\mathbf{H} \parallel [0\,0\,1]$. Notation is the same as in Fig. 2.

Interpretation of the aforementioned feature in the behavior of MnSi is worthy of special consideration. We think that Maleev et al. gave an adequate description of this anomaly in 2006 [4, 5]. Both theoretical analysis [4] and experimental investigation by polarized neutron diffraction [5] showed that this feature is related to reorientation of magnetic domains and the formation of a single-domain magnetic structure at a certain value of the applied magnetic field. In this case, a conical distortion of the helical structure in a field parallel to the helix axis takes place in an arbitrarily small finite magnetic field [4].

Nevertheless, in a number of earlier and recent publications (see, e.g., [2, 6–17]), the feature under consideration has been attributed not only to the reorientation of magnetic domains, but also to a transition from the helical to conical magnetic structure. We think that this treatment is incorrect.

The proposed interpretation originates from the first magnetic phase diagram of MnSi plotted by Kusaka et al. [6]. Although the $H_r(T)$ line corresponding to the anomaly under consideration was not depicted in the phase diagram presented in [6], the proposed interpretation was explicitly used in the text. The magnetic phase diagram of MnSi was somewhat modified in subsequent publications of that group, but the interpretation remained essentially the same (see, e.g., [7]). Later, another group of researchers (see, e.g., [8]) also pointed out that this anomaly is related to the reorientation of magnetic domains, while the conical phase was assumed to form in magnetic fields above the reorientation threshold.

The "second wave" of investigations that used the aforementioned incorrect interpretation of this anomaly was initiated by Thessieu et al. [9]. In the proposed magnetic phase diagram [9], the $H_r(T)$ line was explicitly drawn and interpreted as the boundary between the helical and conical phases. It should be noted that the undistorted helical behavior in the fields below $H_r(T)$ was only described in [9] as hypothetical. However, in subsequent publications, some researchers from the same group (see, e.g., [10–14]) replaced the assumption made in [9] concerning the absence of distortions of helices in the fields below $H_r(T)$ with a rigorous statement of the same without any justification, which they used to describe the obtained results. This interpretation has also been accepted by some other authors (see, e.g., [15–17]) and used up to now, despite the existence of a correct description [4, 5] of the physics of the anomalous behavior of MnSi.

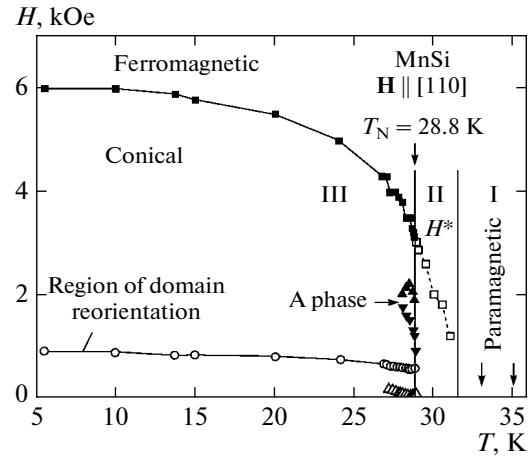

**Fig. 4.** Magnetic phase diagram of MnSi for **H** ∥ [1 1 0]. Notation is the same as in Fig. 2.

What is wrong with this interpretation rather widely used in the literature [6–17]? In fact, this interpretation clearly contradicts the well-known linear field dependence of MnSi magnetization observed starting with the smallest $H$ values (see, e.g., early data [18] on MnSi magnetization). If the helical structure were retained in small fields without conical distortions, the sample magnetization would be zero (or very close to this at finite temperatures) up to the field of conical structure formation—an obvious discrepancy with experiment. In particular, the $M(H)$ curves obtained in our experiments are close to linear starting with the smallest fields—at least with 20 Oe, which is significantly lower than the threshold field of domains reorientation. This behavior was observed in the entire temperature range studied for all three directions of the applied field, in agreement with early data [18] and all other works on the MnSi magnetization. Thus, the aforementioned alternative interpretation is incorrect and the magnetic phase diagrams of MnSi proposed on this basis are inaccurate.

The physical picture of the observed phenomenon can be characterized based on results [4, 5] as follows. An undistorted helical structure is only manifested in a zero field, whereas any finite field directed along the helix axis leads to inclination of the magnetic moments toward the field and the formation of a conical (oblique helical) structure, which in turn results in the appearance of finite magnetization in the field direction. It should be taken into account that four types of magnetic domains exist in a zero field in which the axes of helicoids are directed along the four spatial diagonals of a cube. For example, if **H** ∥ [1 1 1], then the field projections both parallel and perpendicular to the helix axes will exist for domains of three types with other orientations. For an arbitrary **H** direction, situations are possible where the field will be perpendicular to helix axes of the domains of some (no more than two) types. For domains of other types, nonzero field projections on their axes will always exist and lead to conical distortions of the corresponding helices. The case of a field perpendicular to the helix axis has been also considered [4]; it has been shown that a magnetization component along the field must appear and the helices also exhibit distortions, albeit of a nonconical type.

Thus, in the general case, as field **H** increases from zero up to a domain reorientation threshold, the magnetic structure of MnSi has a multidomain character and the domains of no less than two types possess cone-type distortions. Above the domain reorientation field $H_r(T)$, which substantially depends on the orientation of **H**, a single-domain conical structure is formed with the helix axis directed toward the field. If a rather strong field has been preliminarily used to transform the sample to a single-domain state and $H$ is then decreased to values below $H_r(T)$, then our results for $dM(H)/dH$ at **H** ∥ [1 1 1] and $T = 5.5$ K (Fig. 1) show that the sample remains in a metastable monodomain state until the field vanishes. For other **H** directions, a decrease in $H$ below $H_r(T)$ leads to transition of the sample to a multidomain state.

In the temperature interval of 27.2 K ≤ $T$ < $T_N$ = 28.8 K (i.e., in the immediate vicinity of $T_N$), the $M(H)$ curves exhibit another two clearly pronounced anomalies (not depicted in Fig. 1). These features can be assigned, in agreement with earlier investigations, to the formation of the so-called $A$ phase near $T_N$, the boundaries of which in the phase diagram significantly depend on the field direction (see Figs. 2–4).

According to some works (see, e.g., [19]), a characteristic feature of the $A$ phase is that its helix axis is perpendicular to **H**. In recent publications (see, e.g., [10]), the results of neutron scattering measurements for the $A$ phase have been interpreted as evidence of the formation of a more complicated magnetic structure (skyrmion lattice). The possibility of skyrmion formation in MnSi (not only in the $A$ phase) and certain other compounds have been actively discussed in recent years. In particular, the formation of a spontaneous skyrmion phase in MnSi near $T_N$ was originally pointed out in [20]. Data on the polarized neutron scattering [21] have been interpreted as indicative of the appearance of a spontaneous skyrmion state at $T > T_N$. On the other hand, it has been suggested [22] that all results on neutron scattering in MnSi, which were treated as evidence for the existence of skyrmions, can be successfully explained without recourse to the skyrmion model.

At temperatures slightly above the transition ($T_N < T ≤ 31$ K), we have found that, in particular, the $M(H)$ curves exhibit additional features, which are clearly manifested on the field dependences of the differential magnetic susceptibility $dM(H)/dH$ (see the anomaly at $H^*$ on the corresponding curve for $T = 29.0$ K in Fig. 1). At the same time, the $M(H)$ and $dM(H)/dH$ curves obtained at $T > 33$ K did not exhibit these features and could be treated as typically paramagnetic.

The plots of the characteristic field $H^*$ of this feature as a function of the temperature in the interval $T_N < T ≤ 31$ K appear as a natural extrapolation of the temperature dependences of the field of the transition to the induced ferromagnetic phase at $T < T_N$ (see Figs. 2–4). Note also that, for $T > T_N$, the magnetic behavior of MnSi is isotropic, which is characteristic of a paramagnetic state. Thus, the phase existing at $T_N < T ≤ 31$ K can be called (following, e.g., [22]) intermediate, since it occurs between the magnetically ordered phase and the region of typical paramagnetic behavior. In this intermediate state, MnSi exhibits both properties typical of a paramagnetic state (magnetic isotropy) and those not inherent in this state that resemble the behavior of a magnetically ordered phase (magnetization anomaly at $H^*(T)$). As the temperature increases, the magnetization feature observed in the intermediate state gradually disappears; that is, a crossover is observed between the intermediate phase and a phase featuring the typical paramagnetic behavior.

It should be noted that, at $T > T_N = 28.8$ K, the traces of features related to the $A$ phase become almost unobservable and can only be recognized in a very narrow (no more than 0.2 K wide) temperature interval above $T_N$ (see the $dM(H)/dH$ curve for $T = 29.0$ K in Fig. 1). According to the model proposed in [20], which was the first to interpret the behavior of MnSi with the aid of skyrmions, the region of skyrmion formation must be approximately symmetric relative to $T_N$ (see [20] and Appendix therein). Strictly speaking, model [20] refers to the state in a zero field; an attempt to construct a skyrmion model for the $A$ phase of MnSi was made in [10]. The results of our investigation show that, even if the $A$ phase exists at $T > T_N$, this temperature interval is very narrow—much narrower than the region of existence of the $A$ phase at $T < T_N$ (see Figs. 2–4). Taking into account that, at $T_N < T ≤ 31$ K, the $H^*(T)$ curve appears as a natural extrapolation of the field of transition to the induced ferromagnetic phase observed at $T < T_N$, it can be concluded that the genesis of the magnetic properties of MnSi in the intermediate region (28.8 = $T_N < T ≤ 31$ K) is related to the region with the conical magnetic phase (for which skyrmion formation was not considered in the literature) rather than to the $A$ phase. The data on neutron scattering [22] evidence that rather unusual helical fluctuations take place in the intermediate region; based on these results, it was suggested [22] to subdivide the intermediate region into two parts characterized by isotropic and anisotropic (near $T_N$) helical fluctuations. Then, we can naturally suggest that the features observed in the intermediate temperature interval in our investigation correspond to a "collapse" of the fluctuating parts of helicoids in the field $H^*(T)$ by analogy with the collapse of a static conical struc-ture during field-induced transition to the ferromagnetic phase. It is nontrivial that static magnetization measurements allow this collapse to be traced in the intermediate region also.

At $T \geq 33$ K, the MnSi crystal exhibited a typical paramagnetic behavior, showing neither specific features on the $M(H)$ and $dM(H)/dH$ curves nor any evidence of anisotropy.

## 4. CONCLUSIONS

Based on the obtained static magnetization data for MnSi, we have plotted magnetic phase diagrams of MnSi for **H** ∥ [1 1 1], [0 0 1], and [1 1 0] and compared them to the diagrams obtained earlier by other methods. The entire temperature range studied can be subdivided into three regions with decreasing $T$ (see Figs. 2–4):

Region I ($T \geq 31.5$ K), in which MnSi shows a typical paramagnetic behavior.

Region II (28.8 K = $T_N \leq T < 31.5$ K). In this intermediate region, MnSi exhibits both the properties typical of a paramagnetic phase and features characteristic of a magnetically ordered phase.

Region III ($T < T_N = 28.8$ K). This is the region of a magnetically ordered state (containing a region of the $A$ phase with a complex magnetic structure, the nature of which is still not completely clear).

In regions I and II, the magnetization is isotropic, while region III shows a magnetic anisotropy for both the $A$ phase boundaries and features related to the reorientation of magnetic domains. At the same time, in accordance with earlier investigations, the position of the line corresponding to transition to a field-induced ferromagnetic state is practically anisotropic.

In region II, the magnetization curves display characteristic features represented by $H^*(T)$ lines in the magnetic phase diagrams. These $H^*(T)$ lines are a natural extrapolation of the temperature dependence of the field of transition from the conical to the induced ferromagnetic phase observed at $T < T_N$. It has been concluded that the MnSi properties in the indicated intermediate region are related to and governed by those of the conical phase (rather than of the $A$ phase).

In region III, the features observed for small fields are related to the reorientation of magnetic domains and gradual formation of a single-domain magnetic structure in the increasing field. The magnetization curves observed for **H** ∥ [1 1 1] at 5.5 K ≤ $T$ ≤ 27.6 K are irreversible. After transition of the sample into a single-domain magnetic state, the subsequent decrease in field leaves the sample in a metastable single-domain state up to $H = 0$. At the same time, the $M(H)$ curves measured for two other directions of the applied field are almost reversible; that is, as the field decreases below the reorientation threshold, the single-domain structure is not retained and the sample passes to a multidomain state.

After we had prepared the manuscript, we learned that Bauer and Pfeiderer [23] reported on investigation of the magnetic properties of MnSi. The experimental results obtained in [23] are generally close to our data. However, the authors of [23] still think (as in their previous publications [9–11]) that the transition from a helical to conical magnetic phase structure in MnSi takes place only when a certain field ($B_{c1}$) is attained, which evidently contradicts their own data on the $M(B)$ dependence in weak fields.